\documentclass[]{aa}
\usepackage{graphicx,txfonts,color}
\usepackage{natbib}
\bibpunct{(}{)}{;}{a}{}{,} 
\setcitestyle{comma}

\begin{document}
\title{The effect of AGN feedback on the X-ray morphologies of clusters -- simulations vs. observations}
\author{Gayoung Chon$^1$, Ewald Puchwein$^2$ \& Hans B{\"o}hringer$^1$}
\offprints{gchon@mpe.mpg.de}
\institute{
$^{1}$Max-Planck-Institut f\"ur extraterrestrische Physik, D-85748 Garching, Germany\\
$^{2}$Institute of Astronomy and Kavli Institute for Cosmology, 
University of Cambridge, Madingley Road, Cambridge CB3 0HA, UK
}
\date{Submitted 16 March 2016}

\abstract{
Clusters of galaxies probe the large-scale distribution of matter and are a useful tool to test 
the cosmological models by constraining cosmic structure growth and the expansion of the Universe. 
It is the scaling relations between mass observables and the true mass of a cluster through which
we obtain the cosmological constraints by comparing to theoretical cluster mass functions. 
These scaling relations are, however, heavily influenced by cluster morphology. 
The presence of the slight tension in recent cosmological constraints on $\Omega_m$ and $\sigma_8$ 
based on the CMB and clusters has boosted the interests in looking for possible sources for 
the discrepancy. 
Therefore we study here the effect of Active Nuclei Galaxy (AGN) feedback as one of the major mechanisms 
modifying the cluster morphology influencing scaling relations. 
It is known that AGN feedback injects energies up to 10$^{62}$ erg into the intracluster medium, 
controls the heating and cooling of a cluster, and re-distributes cold gas from the centre to outer radii. 
We have also learned that cluster simulations with AGN feedback can reproduce observed cluster 
properties, for example, the X-ray luminosity, temperature and cooling rate at the centre better 
than without the AGN feedback. 
In this paper using cosmological hydrodynamical simulations we investigate how the AGN feedback changes 
the X-ray morphology of the simulated systems, and compare to the observed REXCESS (Representative 
XMM-Newton Cluster Structure Survey) clusters. 
We apply two substructure measures, centre shifts ($w$) and power ratios (e.g., $P_3/P_0$), 
to characterise the cluster morphology, and find that our simulated clusters are more substructured 
than the observed ones based on the values of $w$ and $P_3/P_0$. 
We also show that the degree of this discrepancy is affected by the inclusion of AGN feedback. 
While the clusters simulated with the AGN feedback are in much better agreement with the 
REXCESS $L_X$-T relation, they are also more substructured, which increases the tension with 
observations. 
When classified as non-relaxed/relaxed according to their $w$ and $P_3/P_0$ values, we find that 
there are no relaxed clusters in the simulations with the AGN feedback. 
This suggests that not only global cluster properties, like $L_X$ and T, and radial profiles 
should be used to compare and to calibrate simulations with observations, but also substructure 
measures like centre shifts and power ratios. 
Finally, we discuss what changes in the simulations might ease the tension with observational 
constraints on these quantities.
}

\keywords{galaxies: clusters, cosmology: observations,  
  X-rays: galaxies: clusters, Methods: numerical, 
  Galaxies: clusters: intracluster medium} 

\authorrunning{Chon et al.}
\titlerunning{The effect of AGN feedback on X-ray clusters}
\maketitle
%

\section{Introduction}

Clusters of galaxies have played an important role in accessing the distribution of Dark Matter 
on scales from Mpc up to hundred Mpc, e.g.~\citet{collins00,chon12_scl,r2_underdensity}, and  
have also been proven to be effective tracers of cosmological evolution through the cluster 
mass function which allows to test cosmological models (see e.g.~\cite{vik09,kravtsov12,r2_cosm}). 
The success of cosmological studies with clusters of galaxies heavily relies on the fact that 
mass observables are a clean probe of the total mass of a system. 
As shown explicitly by~\cite{r2_cosm} the scaling relation that connects the measured X-ray luminosity 
to the total mass has a large influence on the cosmological constraints, most sensitively through 
$\sigma_8$ and $\Omega_m$. 
Hence it is crucial to understand scaling relations and their statistical properties as accurately
as possible.

In our previous study we showed that the dynamical states or equivalently morphologies of clusters
help us to understand the scatter introduced in scaling relations~\citep{chon12}. 
For instance we showed that the normalisation of scaling relations is systematically higher or 
lower depending on the dynamical state of a cluster.
This is very interesting in light of an apparent tension of the constraints on $\sigma_8$ and $\Omega_m$ 
from the analysis of the comic microwave background (CMB) and cluster 
counts~\citep{planck15_cluster,planck15_cosm,hasselfield13,vik09,r2_cosm} because it is this normalisation 
parameter of the scaling relation that 
may be able to ease the tension in addition to the much-discussed hydrostatic mass bias. 

One of the major mechanisms that modify the distribution of the intra-cluster medium (ICM) is 
Active Galactic Nuclei (AGN) feedback. 
The lack of highly cooling gas at the centres of clusters in X-ray observations provides a direct 
evidence that there is a heating mechanism which prevents the ICM from 
over-cooling~\citep{peterson01,boehringer02}.
The most viable source for this heating is provided by central black holes injecting energy into the ICM.
There are two known modes of AGN feedback, radio and quasar modes, depending on the accretion rates of
central black holes (see e.g. the reviews by~\cite{mcnamara07,mcnamara12}).
The quasar mode is effective in the early phase of the AGN evolution, when the black hole mass 
accretion rate is high and most of the feedback energy is emitted as radiation.
For clusters of galaxies the most relevant mechanism is the radio mode in operation at low black hole
mass accretion rates.
In this mode the feedback energy is released primarily as mechanical power which heats the atmosphere
through buoyant bubbles rising from the centre initiated by energetic AGN jet events.
Also the heating by shock waves triggered by the AGN jets has been observed in a few cases.

\cite{springel05_bh} developed techniques to incorporate the AGN feedback in cosmological hydrodynamics 
simulations, and modifications were made, for example, by~\cite{sijacki06,sijacki07} 
and~\cite{fabjan10} which included both the quasar and radio feedback modes. 
There are several studies which investigated the effect of AGN feedback based on the X-ray properties
of clusters and groups including scaling relation, temperature, and metal abundance profiles
(see e.g.~\citet{puchwein08,fabjan10,mc10,plane14}). 
However, none of these studies directly investigated its effect on the morphologies of clusters
and compared to observations, which is the main aim of this paper.

In a previous study \cite{puchwein08} showed that including AGN feedback in the simulations 
of the formation of galaxy clusters reproduces some of the important scaling relations of 
cluster parameters from X-ray observations, that are not reproduced without the effect of AGN feedback. 
In particular the relation of the X-ray luminosity and the intracluster plasma temperature as well as 
the intracluster gas mass fraction were well matched to observations for the simulations including 
AGN feedback. 
Here we take this simulation result, and explore the effect of the AGN feedback on the morphology of X-ray 
clusters.

We employ two common substructure measures, the power ratios and centre shifts, which are well
tested for the X-ray observations and simulations (see, e.g.~\cite{rexcess_sub,chon12,mahdavi13,rasia13}) 
to determine quantitatively the degree of substructure, and compare the results from simulations with
and without AGN feedback to X-ray observations.
For the X-ray observations we consider the REXCESS clusters.
They form a statistical sample which is closer to a volume-limited sample drawn from the ROSAT-ESO 
Flux Limited X-ray galaxy cluster Survey (REFLEX)~\citep{reflex1,reflex2-1,reflex2-2},
all of which have deep exposures with XMM-Newton observations. 
The REXCESS clusters have been scrutinised for X-ray scaling relations and 
morphologies~\citep{pratt09,rexcess_sub}, hence provide a good observational basis to compare 
to simulations.

The paper is structured as follows. 
In section 2 we describe the simulation data and the observational data, and provide a brief 
summary of the substructure measures in section 3.
In section 4 we study the AGN feedback effect on the cluster morphology in the simulations, 
and compare these results to the REXCESS. 
We conclude the paper with a summary in section 5.

\section{Sample description}

\subsection{The simulated cluster sample}
\label{sec:sim_sample}

We use the galaxy cluster and group sample from~\cite{puchwein08}, which consists of 21 re-simulations 
of Millennium simulation \citep{springel05} halos performed with the zoom-in technique. 
The same flat $\Lambda$CDM cosmology as in the parent Millennium simulation was adopted, 
$\Omega_m$=0.25, $n_s$=1, $\sigma_8$=0.9, and $\Omega_b$=0.04136. 
The resolution of the re-simulated halos with virial masses below $2\times10^{14} h^{-1} M_\odot$ is:  
the DM particle mass is $m_{\rm DM} = 3.1 \times 10^7 h^{-1} M_\odot$, the gas particle mass is
$m_{\rm gas} = 6.2 \times 10^6 h^{-1} M_\odot$ and the physical softening is $\epsilon = 2.5
\, h^{-1} \rm kpc$.
For the four most massive clusters a lower resolution was used with a DM particle mass 
of $m_{\rm DM} = 1.1 \times 10^8 h^{-1} M_\odot$ and a gas particle mass of $m_{\rm gas} = 2.1
\times 10^7 h^{-1} M_\odot$.
All re-simulations include hydrodynamics, radiative cooling assuming a primordial gas composition, 
heating by an external UV background, star formation and supernovae feedback.
The selection of this sample was based only on halo mass covering a large range from 
$M_{200} = 8 \times \! 10^{12} \mathrm{M}_{\odot}$ to $1.5 \times \! 10^{15} \mathrm{M}_{\odot}$. 
Here and throughout this work the densities quoted in the subscript indices of spherical overdensity 
masses and radii are given in units of the critical density of the universe at the cluster redshift.
The simulations employ a "traditional" entropy-conserving formulation of smoothed particle 
hydrodynamics (SPH)~\citep{springel02}. 
No artificial conduction/mixing prescription is used in this scheme. 
The runs were carried out with the P-Gadget3 code (last described in~\cite{springel05}).
For each halo two kinds of re-simulations are available. 
They were performed either with or without a model for the growth of supermassive black 
holes (BHs) and associated feedback processes as in~\cite{sijacki07} ~\citep[also see][]{springel05_bh}. 
The stellar components of these clusters were studied in \citet{puchwein10}, while their 
lensing properties were investigated in \citet{mead10}.

To incorporate BH growth and feedback it was assumed that any halo above a mass of 
$5 \! \times \! 10^{10} h^{-1} M_\odot$ contains a seed BH with a mass of $10^5 h^{-1} M_\odot$. 
The BHs are allowed to grow by mergers with other BHs and by accretion of gas. 
The prescription of the latter is motivated by the Bondi-Hoyle-Lyttleton model,
but with the Eddington limit additionally imposed.

Two modes of AGN feedback were adopted depending on the BH accretion rate, above 0.01 of 
the Eddington rate the quasar mode, and below the radio mode. 
In the quasar mode where feedback is assumed to be predominantly radiative, a low coupling 
efficiency of 0.5\% of the rest mass energy of the accreted gas is used. 
This energy is continuously injected in the form of thermal energy into nearby gas particles. 
In the radio mode, feedback is assumed to happen by recurrently inflating AGN-heated bubbles 
into the ICM.
We here employ a larger mechanical coupling efficiency of 2\% of the accreted rest mass energy,
which is in good agreement with X-ray observations of elliptical galaxies \citep{allen06}. 
The radio mode feedback energy is injected, i.e. a bubble event is triggered, whenever the mass 
of the BH has grown by a factor $\Delta M_{\it BH} / M_{\it BH} \geq 10^{-4}$. 
The accumulated feedback energy is then injected thermally into a spherical bubble with 
a radius given by $R_{\rm bub} = R_{\rm bub,0} [(E_{\rm bub} / E_{\rm bub,0}) \, / \, (\rho_{\rm bub} / 
\rho_{\rm bub,0})]^{1/5}$, where the default bubble radius $R_{\rm bub,0} = 30 \, h^{-1} {\rm kpc}$ 
at a bubble energy of $E_{\rm bub,0} = 10^{55} {\rm erg}$ and an ICM density of 
$\rho_{\rm bub,0} = 10^{13} \, h^{-2} M_\odot {\rm Mpc}^{-3}$ fixes the overall normalization of 
the radii of bubbles.
The scaling with the bubble energy, $E_{\rm bub}$, and the ICM density at the bubble position, 
$\rho_{\rm bub}$, is motivated by solutions of the expansion of radio cocoons 
\citep[e.g.][]{heinz98} and ensures that more energetic feedback events result in larger bubbles, 
while a denser ICM confines the bubble size more strongly. 
The centre of the bubble is chosen randomly within a radius of $5/3 \times R_{\rm bub}$ around the BH.
The adopted BH model leads to a self-regulated BH growth as shown in~\cite{sijacki07}.

\subsection{X-ray properties of simulated and observed clusters}

Two sets of 21 clusters, with and without AGN feedback, were used to create X-ray surface 
brightness images in three projections resulting in 126 clusters images. 
The cluster mass range described above corresponds to X-ray spectroscopic temperatures between 
0.4 to 7.4~keV.
A constant metallicity of 0.3$Z_\odot$ was assumed in the synthetic X-ray analysis.
The X-ray images were produced in the [0.5-2.0]~keV band.

For the comparison with observations we used the representative XMM-Newton cluster structure survey
(REXCESS) data \citep{rexcess} comprised of 31 galaxy clusters which are selected to be morphologically 
representative \citep{rexcess_sub}. 
The cluster selection is closer to volume-limited than flux-limited, which is important to note, as these 
selections do affect the distributions of the morphologies of clusters (Chon et al., in prep.).
The lower limit of the temperature in this sample is 2~keV. 

While our results will be shown for all of the simulated clusters, we will make a closer comparison 
to the observations with the four simulated clusters that fall into the same temperature or mass range 
as the REXCESS sample.

\section{Structural analysis}

To characterise the degree of substructure we concentrate here on two methods:
the power ratios~\citep{buote95} and centre shifts (e.g.~\citet{mohr93,poole06}). 
We use the same technique as in~\cite{rexcess_sub} with modifications 
as descried in~\cite{chon12} such that the final classification of the morphology reflects 
the mean dynamics over the entire cluster out to $r_{500}$.

\subsection{Power ratio calculation}

The power ratio method first introduced by~\cite{buote95} was motivated by the assumption that
the X-ray surface brightness closely traces the projected two dimensional mass distribution
of a cluster. 
A multipole decomposition of such a projected mass distribution provides moments which are 
identified as power ratios after normalisation by the zeroth moment.
In practice the power ratio analysis is applied to the surface brightness distribution.

The moments $P_m$ are defined as
\begin{equation}
P_0 = \left[ a_0 \ln (R_{\rm ap}) \right]^2
\end{equation}
\begin{equation}
P_m = \frac{1}{2 m^2 R_{\rm ap}^{2m} } \left( a_m^2 + b_m^2 \right)
\end{equation}
\noindent where $R_{\rm ap}$ is the aperture radius in units of $r_{500}$. 
The moments $a_m$ and $b_m$ are calculated by:
\begin{equation}
a_m(r) = \int_{r \le R_{\rm ap}} d\vec{x} ~S(\vec{x}) ~r^m \cos (m\phi)
\end{equation}
\noindent and
\begin{equation}
b_m(r) = \int_{r \le R_{\rm ap}} d\vec{x} ~S(\vec{x}) ~r^m \sin (m\phi), 
\end{equation}
where $S(\vec{x})$ is the X-ray surface brightness image, and the integral extends over all pixels
inside the aperture radius. 
$a_0$ in Eq. (1) is thus the total radiation intensity inside the aperture radius.

Since all $P_m$ are proportional to the total intensity of the X-ray image, all moments are normalised 
by $P_0$ resulting in the so-called power ratios, $P_m/P_0$.
For brevity we refer to $P_m/P_0$ as $P_m$ in the rest of the paper. 

Similarly to all previous studies, we only make use of the lowest moments from $P_2$ to $P_4$. 
Before the multipole moments are determined, the centre for the calculations is found 
by determining the centre of mass in the vignetting and gap corrected surface brightness 
images. 
The dipole, $P_1$, should therefore vanish, which is checked during the calculations. 
$P_2$ describes the quadrupole of the surface brightness distribution, which is not necessarily 
a measure of substructure. 
In practice, low to moderate values of $P_2$ are found for regular elliptical clusters, 
while larger values of $P_2$ are a sign of cluster mergers.
The lowest power ratio moment providing a clear substructure measure is thus $P_3$. 
$P_4$ describes substructure on slightly finer scales and is found to be correlated with 
$P_2$ here and in previous studies~\citep{chon12}.

Typically $P_m$($r$) is evaluated at $r_{500}$. 
Due to the $r^m$ weighting in Eqs. (3) and (4), it is then predominantly influenced 
by the dynamical state in the outskirts of a cluster. 
To overcome this strong bias to the outer part of the cluster, we measure the averaged 
power ratios at ten radii defined by fractions of $r_{500}$.
Further detailed discussions and comparisons to other definitions are found in~\cite{chon12}.

In the case of observed clusters the uncertainty of the power ratio measurement and the influence of 
photon noise are studied by simulations in which an additional Poisson-noise is imposed on the count 
images with background. 
This ``Poissonisation'' is equivalent to the Poisson-noise introduced by the observation involving a finite 
number of detected photons. 
We interpret the variance of the power ratio results from the simulations as the measurement uncertainty. 
We then make the assumption that the additional power introduced by the Poissonisation is similar to the extra 
power in the power ratios introduced by the photon shot noise of the observation. 
We therefore subtract the additional noise found in the mean of all simulations compared to the observations 
from the observational result.

\subsection{Centre shifts}

The centre shift measures the stability of the X-ray centre calculated at different radii, and is formulated 
as~\citep{poole06}:
\begin{equation}
w~ =~  \left[ \frac{1}{N-1}~ \sum \left( \Delta_i -  \langle 
\Delta \rangle \right)^2 \right] ^{1/2} ~\times~ \frac{1}{r_{500}} 
\end{equation}
\noindent where $\Delta_i$ is the distance between the mean centroid and the centroid of the $i$th aperture. 

The centroid of each aperture is found by determining the ``center of mass'' of the photon distribution in each 
aperture, which was already used for the centering prior to the power ratio determination. 
The resulting $w$ is then the standard deviation of the different centre shifts (in units of $r_{500}$).
We use the mean centroid value of all apertures as the reference centre.

For the X-ray observations of clusters the uncertainties in the centre shifts and in the $w$ parameter are 
determined with the same simulations as the uncertainties of the power ratios, i.e., by using Poissonised 
re-sampled cluster X-ray images. 
The standard deviation of the $w$ parameter in the simulation is used as an estimate of the measurement 
uncertainties. 
We do not use the noise-bias subtracted $w$-parameter as in the case of the power ratios since the bias 
correction is mostly much smaller than the errors and the bias correction does not shift the $w$-parameter 
to alter the classification of the cluster morphology.

We performed end-to-end Monte-Carlo simulations of the Poissonised data analysed exactly the same way 
as we calculate the power ratios and centre shifts with the original X-ray data. 
This ensures that, for example, the systematics introduced by the photon shot noise is properly taken 
into account in the parameter uncertainties.
We note that those simulated clusters in the same mass or equivalently temperature range as those of the 
REXCESS data are represented by larger symbols in the figures where this distinction is necessary.

\subsection{Variations of observed substructure due to projections}

The line-of-sight projection of a cluster leaves the properties related to the third dimension unconstrained. 
This ambiguity introduces scatter in the morphology parameters for an individual cluster.
The degree of scatter can be studied with the substructure measures from three different projections of the 
same cluster. 

\begin{figure}
  \includegraphics[width=\columnwidth]{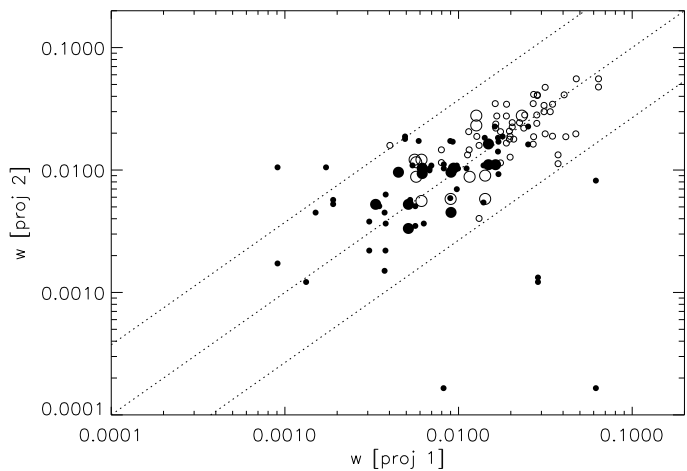}
  \caption{
    Comparison of the results of the centre shift analysis for two projections of the same cluster. 
    Filled symbols show the clusters without AGN feedback and open symbols those with the feedback. 
    Larger circles represent those clusters above 2~keV as in the REXCESS sample. 
    The central dotted line marks an one-to-one relation, enclosed by three sigma scatter.
  }
  \label{fig:proj-w}
\end{figure}

\begin{figure}
  \includegraphics[width=\columnwidth]{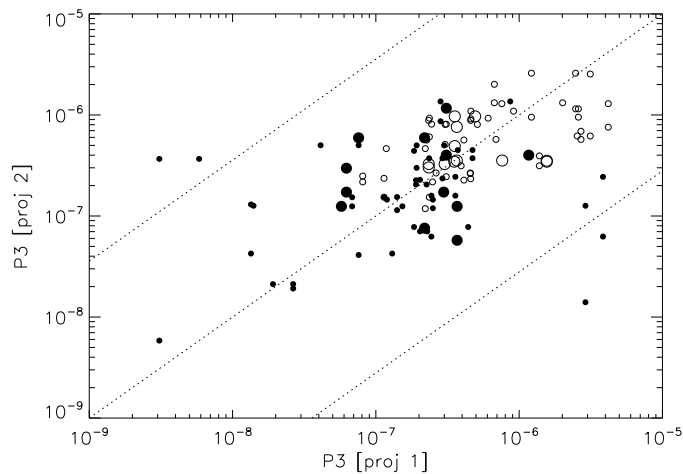}
  \caption{
    Comparison of the power ratio, P3, for two projections of the same cluster. 
    Symbols are defined as those in Fig.~\ref{fig:proj-w}.
}
\label{fig:proj-p3}
\end{figure}

We show the results for the power ratio, $P_3$, and for the centre shift parameter in Figs.~\ref{fig:proj-w} 
and~\ref{fig:proj-p3} for all 126 cases.
Both figures show a log-log correlation between different projections with the three sigma scatter
drawn as dotted lines around a central dotted line denoting an one-to-one relation.
The clusters with AGN feedback (open circles) have a larger value of $w$ and a smaller scatter 
than those without (filled). 
The scatter is smaller for more massive (large circles) than for the less massive clusters 
at least in the runs without AGN. 
There are nine clusters whose $w$ value in one projection is significantly larger than in the other
one, which are located outside the three sigma lines. 
It is interesting to note that they are on the least massive end of the mass range in the simulations. 
Eight of those are less massive clusters without AGN feedback, and one with AGN feedback.
We remark that on average the scatter of $w$ is smaller than that of $P_3$, which may indicate that 
$w$ has a stronger constraining power of substructure than $P_3$ at least for a moderately sized
sample despite those five clusters with very large departures from the one-to-one relation in $w$.

The influence of AGN feedback will be investigated in more detail in the next section.

\section{The effect of AGN feedback}

\subsection{Simulation}

\begin{figure}
  \includegraphics[width=\columnwidth]{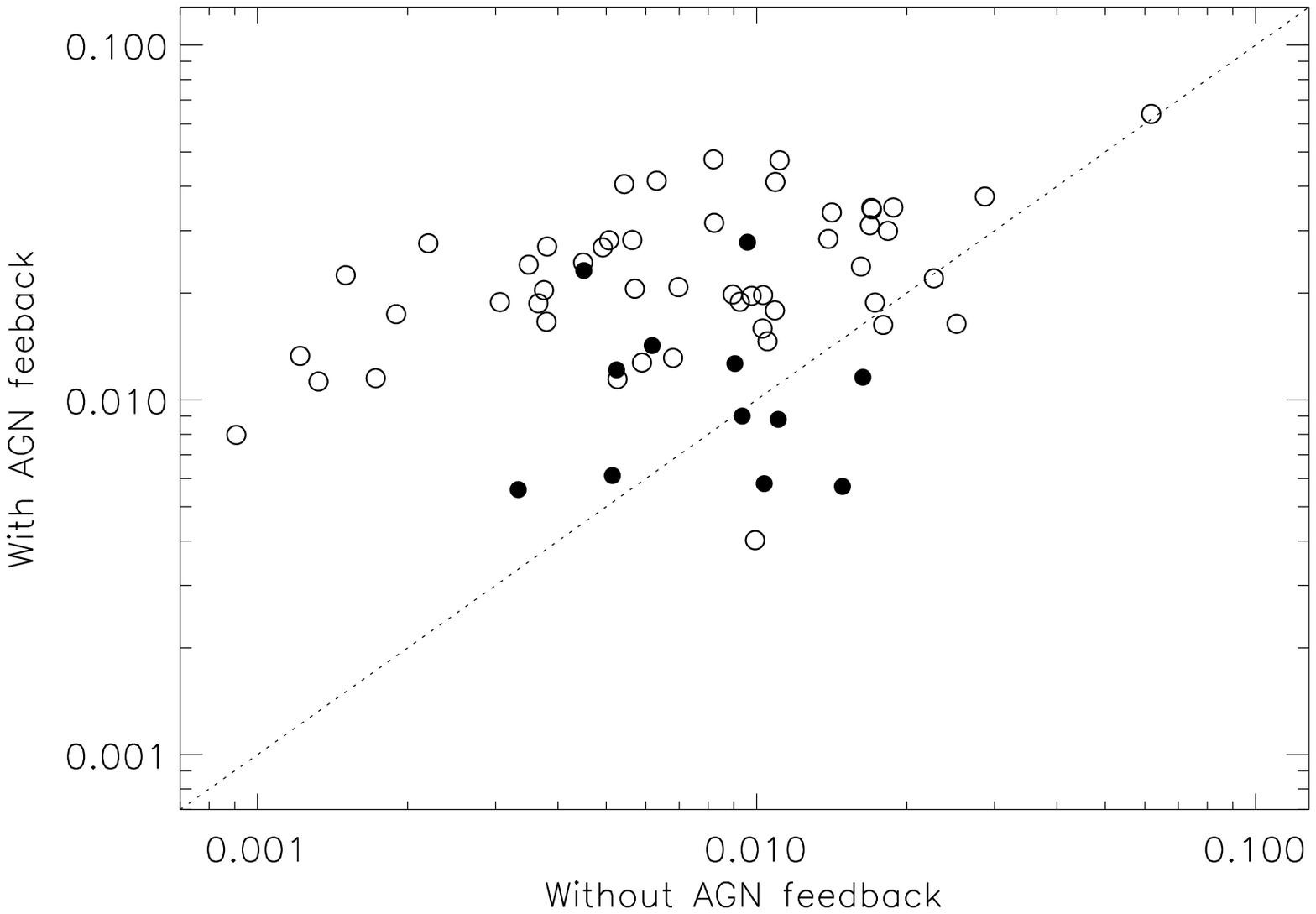}
  \includegraphics[width=\columnwidth]{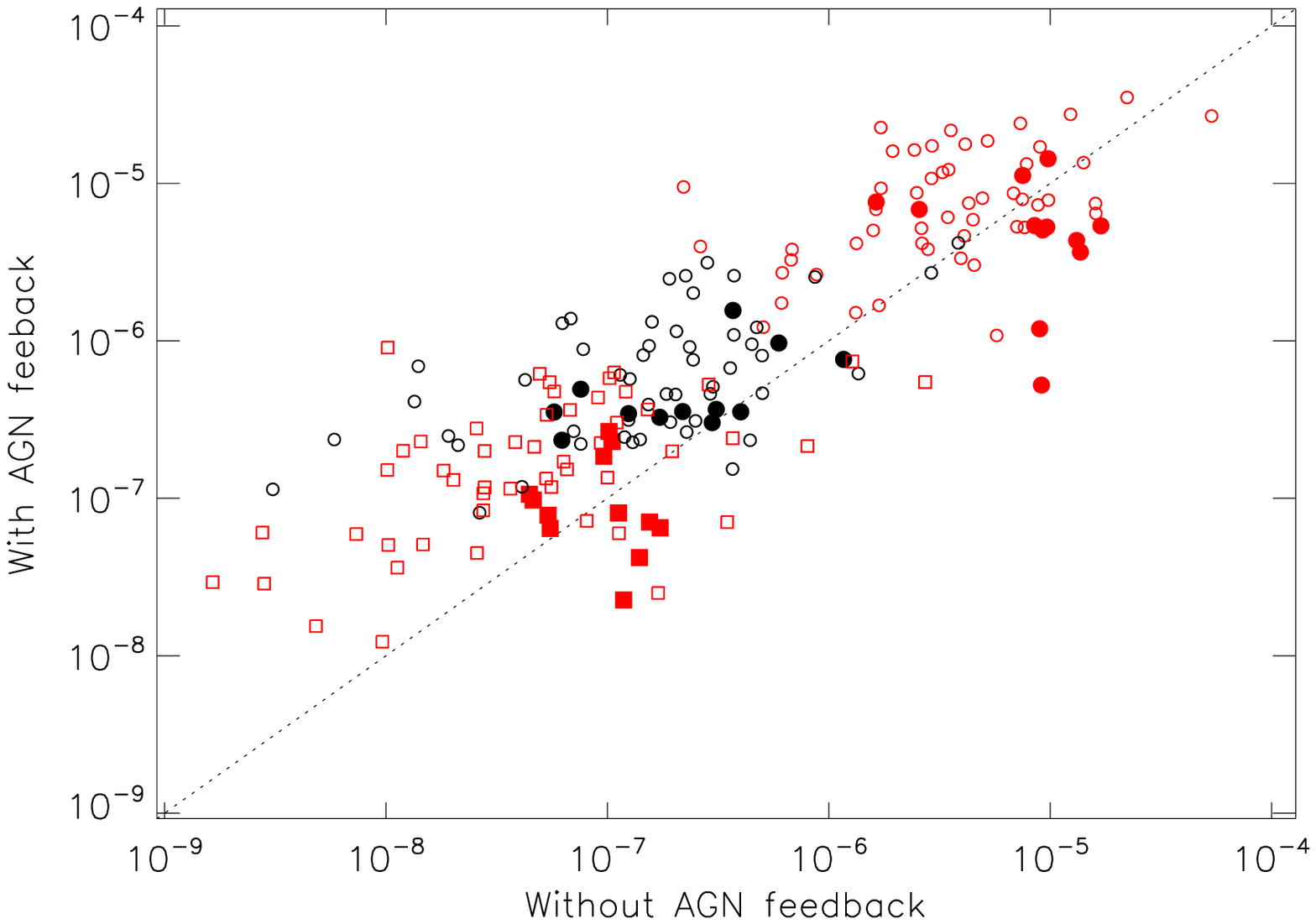}
  \caption{
    (Top) 
    Comparison of the centre shift values.    
    (Bottom)
    Comparison of the power ratios, $P_2$ (red circles), $P_3$ (black circles) and 
    $P_4$ (red squares).
    In both panels massive clusters (filled) and low mass ones (open circles) 
    are shown together with an equality line (dashed).
   }
   \label{fig:ep_bhobh}
\end{figure}

Figs.~\ref{fig:proj-w} and~\ref{fig:proj-p3} indicate that AGN feedback creates more 
substructure in the X-ray surface brightness maps, and we study here the degree of how 
much more substructure is introduced by the AGN feedback.
Fig.~\ref{fig:ep_bhobh} compares directly the values of $w$ (top panel) and $P_2$, $P_3$ and $P_4$ (bottom) 
for simulations with and without AGN feedback for the same cluster in the same projection. 

The upper panel clearly shows that low mass objects (open circles) typically have a larger $w$ value 
in runs with AGN feedback than in runs without.
For massive clusters (filled circles) this offset becomes less significant. 
This finding is not too surprising as the energy input of the AGN pushes the gas out of the very central
region and this has a larger effect on the global scale in low mass groups with shallower 
gravitational potentials compared to massive clusters where the deeper gravitational potential
can confine the feedback effect more to the central region. 

In Table 1 we list the mean ratios of the centre shifts and power ratios between the runs
with and without AGN feedback. 
We find that the mean ratio of the $w$ values between with and without the AGN feedback is 
approximately six times larger in less massive clusters than that in massive clusters.
More concretely the mean value of $w$ with the AGN feedback is 0.012 for the massive clusters 
and 0.025 for the less massive ones.
This suggests that the current implementation of the AGN feedback not only pushes the gas out to 
larger radii in less massive clusters but also produces a more asymmetric gas distribution. 
This larger effect in low mass objects is consistent with previous simulation results, and 
was found to be required to match observed cluster scaling relations and gas fractions 
at the low mass end~\citep{puchwein08}.

We also show a similar trend for all three power ratios in the bottom panel of Fig.~\ref{fig:ep_bhobh}. 
The bulk mean amplitude of power ratios increases from $P_4$ to $P_2$. 
Overall the power ratios that are more heavily influenced by the inclusion of the AGN feedback are $P_3$ 
and $P_4$.
For $P_2$ the less massive clusters have a mean ratio of 3.8 between the runs with and without 
AGN feedback while that of the massive clusters is 1.1 as shown in Table 1.  
However, the effect of the AGN feedback is more pronounced in $P_3$ and $P_4$ as they measure the 
distortion of gas distribution on finer scales. 
\begin{table}
\begin{center}
\centering
\caption{Ratios of the substructure parameters between the runs 
with and without AGN feedback divided into two mass groups. 
This division is done at 2~keV, which is the lower temperature
limit of the REXCESS cluster sample. 
This roughly corresponds to a mass cut of 8$\times 10^{13}$ M$_\sun$/h.
}
\begin{tabular}{c c c c c}
\hline
\multicolumn{1}{c}{} & 
\multicolumn{1}{c}{$w$} & 
\multicolumn{1}{c}{$P_2$} &
\multicolumn{1}{c}{$P_3$} &
\multicolumn{1}{c}{$P_4$} \\
\hline
light & 10.66  &    3.80   &   7.69   &   7.01 \\
massive &  1.69  &    1.09   &   2.69  &    1.32 \\
\hline
\end{tabular}
\end{center}
\label{tab:tab1}
\end{table}
As was seen in the centre shifts we also find that the ratio is lower for massive clusters 
than the less massive ones, typically by a factor of three to five.
The mean value of $P_3$ with AGN feedback is 5.34$\times 10^{-7}$ for the massive clusters
and 9.10$\times 10^{-7}$ for the less massive ones. 

One primarily expects that AGNs heat the intracluster medium, to prevent a massive cooling 
and condensation of gas in the centre, to disturb the distribution of the central gas, and 
to puff the cold gas out to a larger distance. 
Hence it is not surprising that clusters with AGN feedback show a larger degree of substructure 
in the gas distribution than those without, and that the gas distribution of less massive systems 
are more affected by AGN feedback than that of massive clusters. 
We will compare in the next section our findings in the simulations with the observations.

\subsection{Comparison to observations}

\begin{figure}
  \includegraphics[width=\columnwidth]{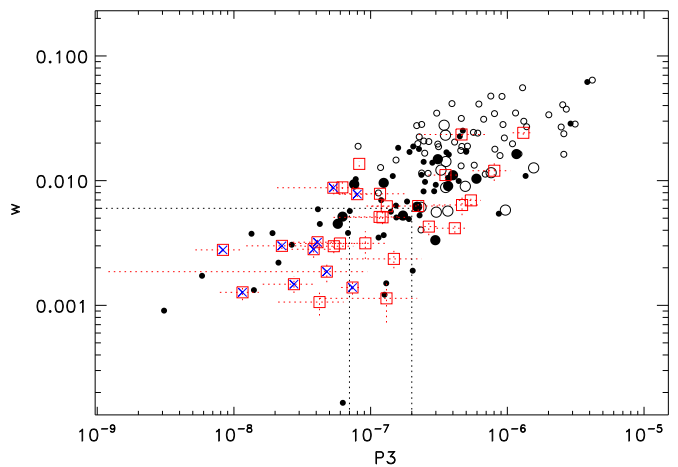}
  \caption{
    Comparison of the centre shift parameter $w$ and power ratio $P_3$ for observations 
    (red squares) and simulations (circles).    
    Open circles represent clusters with and filled ones without AGN feedback.
    The massive clusters are marked with larger circles. 
    Among the observations cool-core clusters are marked by blue crosses. 
    Dotted lines give guidelines to distinguish different morphological types (see text
    for details). 
  }
  \label{fig:ep_so_wp3}
\end{figure}

We use the REXCESS clusters~\citep{rexcess} to compare the simulations to observations.
In Fig.~\ref{fig:ep_so_wp3} we compare the $w$ and $P_3$ values for the simulations (black circles) and 
observations (red squares with errors). 
Overall there is a correlation between the two parameters with some scatter. 
A large fraction of the simulated clusters are located in the upper right corner of the
$w$-$P_3$ plane suggesting more substructures. 
Among the simulated sample the clusters with the AGN feedback (open circles) are more substructured
than those without (filled). 
Also the less massive clusters and groups (small circles) tend to be at the lower left or upper right corner 
of the figure displaying a larger range of cluster morphologies than for the massive clusters (large circles)
which are distributed in the middle of the plot.
Those REXCESS clusters identified as cool-core clusters \citep{pratt09} are shown with the extra
blue crosses. 
A fair comparison of the observations and simulations needs to be restricted to the same mass range 
(indicated by large circles for the simulations).
This comparison would then bring the observations and the simulations closer in the $w$-$P_3$ plane
albeit retaining the tendency that the simulated clusters have still larger $w$ and $P_3$ 
parameters.
The ratio of the mean $w$ parameters between the simulations without AGN feedback and 
the observations is 1.4 and that of $P_3$ is 1.6.
These ratios increase to 1.9 for $w$ and 2.6 for $P_3$ if we consider instead the runs with
AGN feedback. 

The three dotted lines in Fig.~\ref{fig:ep_so_wp3} are defined and used in~\cite{chon12} to classify
the morphology of a cluster in a more qualitative way, which also proved useful for other purposes. 
This classification divides clusters into three groups, the disturbed ($w$>6$\times 10^{-3}$ or 
$P_3$>2$\times 10^{-7}$), relaxed ($w$<6$\times 10^{-3}$ and $P_3$<6$\times 10^{-8}$) and
intermediate, which occupy the bottom narrow strip in Fig.~\ref{fig:ep_so_wp3}. 
The REXCESS sample then contains 15 disturbed, 6 intermediate and 10 relaxed clusters. 
As was found in~\cite{chon12} this division of morphologies divides the sample into two 
similarly sized groups, disturbed and less disturbed. 
In contrast 75\% of the massive clusters simulated without AGN and all of the massive clusters 
simulated with AGN are disturbed. 
Hence we find a significant discrepancy in morphologies between the simulated and observed clusters, 
which gets larger when AGN feedback is included. 
This indicates that some relevant physics might be missing or is incorrectly treated in the simulations. 
Furthermore, the increased discrepancy in morphology in runs with AGN feedback 
might suggest, that the way AGN feedback was introduced in the simulations was perhaps 
more violent than in nature and in particular introduced more asymmetric disturbances. 
If the energy deposition would preserve more of the symmetry, the morphology would be 
less distorted with the same amount of feedback energy input.

\begin{figure}
  \includegraphics[width=\columnwidth]{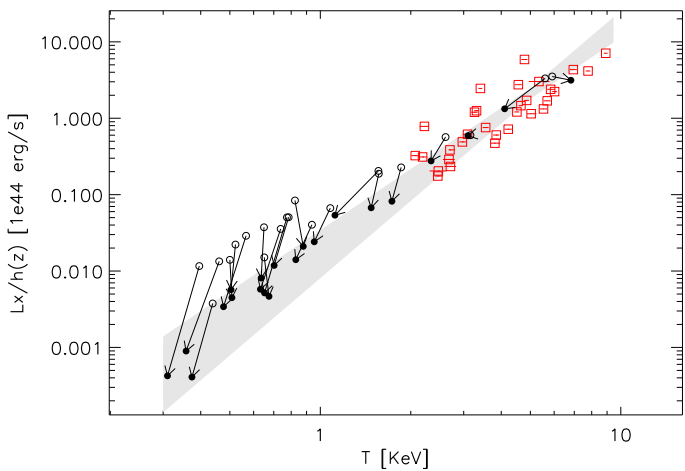}
   \caption{
     X-ray luminosity as a function of the spectroscopic temperature for observations (red squares)
     and simulations.
     Arrows mark the change of the cluster properties starting from the case without the AGN (open
     circles) to with the AGN (filled circles).
     The shaded ares shows the 1$\sigma$ uncertainty of the best-fit REXCESS $L_x$-T scaling relation,
     which was derived for clusters above 2~keV.
   }
   \label{fig:ep_so_lxt}
\end{figure}

This finding does not imply a complete revision of the treatment of AGN feedback in 
cosmological simulations for clusters. 
Fig. 2 of~\cite{puchwein08} showed that there is a very good agreement between observations 
and simulations for the $L_X$-T scaling relation. 
We compare here the $L_X$-T relation for the simulations and for REXCESS in Fig.~\ref{fig:ep_so_lxt}.
The REXCESS data points in red squares were used to obtain the best-fit scaling relation 
in~\cite{pratt09} shown in the gray region. 
We extend this region down to the lower temperature in the plot just for a comparison 
to the simulated clusters. 
It is clear that the simulated clusters with AGN feedback provide a much better agreement with 
the observed scaling relation.  

\begin{figure}
  \includegraphics[width=\columnwidth]{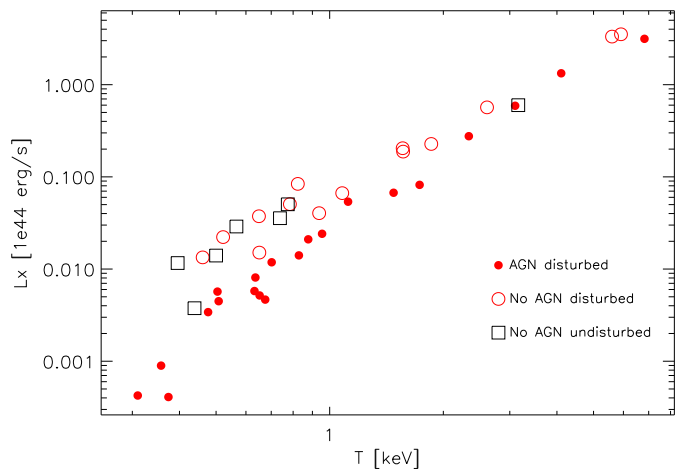}
   \caption{
     X-ray luminosity as a function of the spectroscopic temperature represented by
     the morphological description.
   }
   \label{fig:ep_morph}
\end{figure}

Finally we examine a possible dependency of the morphological types on the physical properties
of clusters in Fig.~\ref{fig:ep_morph}. 
The cluster morphology classification was based on the results of Fig.~\ref{fig:ep_so_wp3}. 
Except for one cluster all other six relaxed clusters are below 2~keV, and there is no other
clear segregation of relaxed and unrelaxed clusters with respect to the $L_X$--T relation, 
as was found in~\cite{chon12} and~\cite{rexcess_sub}.

Hence we conclude in this section with a statement that the simulated clusters have more 
substructures than the observations as measured by $w$ and $P_3$, and the degree of 
substructures for the simulation with the AGN feedback is much higher than the one 
without despite the fact that the observed scaling relation, $L_X$--T, is better 
reproduced with the AGN feedback.
This demonstrates that the substructure measures, $w$ and $P_3$, are sensitive to the included 
feedback physics and suggests that they should be used in addition to global properties, 
like $L_X$ and T, and radial profiles to test cluster simulations against observations.

Finally, we want to reflect on what changes in the simulated physics might ease the tension 
with observations. 
In fact some of the most recent simulations of galaxy cluster formation do get a mixture of 
cool-core and non-cool-core clusters \citep[][]{hahn15, rasia15} and might potentially
also be doing better in terms of the X-ray substructure measures employed here. 
Both works differ from our simulations in the treatment of the hydrodynamics.
\citet{hahn15} use the Eulerian adaptive mesh refinement code {\sc ramses}. 
\citet{rasia15} employ a more modern SPH version with an updated interpolating kernel
and artificial conduction, which allows to better resolve fluid instabilities and results 
in more mixing \citep{beck16}. 
While these two hydro algorithms are quite different, they result in almost identical 
thermodynamic profiles in a non-radiative galaxy cluster simulation in which the effect 
of the hydrodynamics can be directly compared \citep{sembolini16}. 
These profiles also differ significantly from those obtained with classic SPH algorithms 
like the one employed in our runs. 
Classic SPH codes also tend to produce somewhat more cold gas clumps due to a suppression 
of fluid instabilities that could otherwise shred them. 
This could potentially increase X-ray substructure measures. 
We think, however, that this is not a large effect and therefore does not play a role
in the comparison of simulations with and without AGN feedback.

Our simulations compute radiative cooling based on a primordial gas composition. 
It would certainly be desirable to replace this with a chemical evolution and 
metal-dependent cooling prescription. 
In the synthetic X-ray analysis an ICM metallicity of 0.3 times the solar value 
was assumed. 
This is in broad agreement with typical observed values. 
Nevertheless, a more realistic inhomogeneous enrichment could somewhat alter 
the X-ray morphology in particular in low mass objects where metal line emission 
often dominates over free-free Bremsstrahlung.

In terms of the AGN feedback implementation \cite{hahn15} and~\cite{rasia15} inject 
the feedback energy in the immediate neighbourhood of the BH particle, while in our 
simulations the injection of bubbles is somewhat off-center to mimic the observed 
positions of X-ray cavities/radio bubbles. 
AGN jets may shoot out to these radii with very limited coupling to the ICM further inside. 
In practice it is, however, often difficult to achieve high enough numerical resolution 
to resolve bubbles of realistic sizes well enough to accurately follow their dynamics. 
One is then left with the choice of either injecting very poorly resolved bubbles or 
making them somewhat larger.
The adopted parameters controlling the bubble size in our simulations 
(see Sec.~\ref{sec:sim_sample}) fall in the latter regime. 
As the cluster-centric radius of the bubble centre is randomly chosen within $5/3$ times 
the bubble radius, this also results in more off-centred bubbles, which could potentially 
disrupt an approximately spherical symmetric gas distribution more strongly and might result 
in larger substructure measures. 
In very high resolution simulations, more accurately matching the bubble properties and 
positions to observations is thus worth exploring.

\section{Summary}

We used two measures of substructures, the centre shifts, $w$, and power ratios, mainly $P_3$,
to diagnose the degree of substructures in X-ray clusters for both simulations and observations. 
The line-of-sight projection of a cluster introduces an ambiguity in the measured degree of
substructures leading to scatter for individual clusters. 
Thus these measures may not lightly be used to judge an individual case, but they are very 
useful in treating statistical samples of clusters. 

We find that a combination of $w$ and $P_3$ provides a reliable measure to diagnose dynamical 
states of clusters as found in~\cite{chon12}.
This pair of parameters was used to divide the cluster samples into three classes of morphologies
or equivalently dynamic states, being disturbed, intermediate and relaxed.
This classification divides the REXCESS clusters into two similarly sized groups, one group of 
the disturbed and another of intermediate and relaxed clusters.
However, the simulated clusters are dominated by clusters with much more substructures for both 
types of simulations, i.e. with or without the AGN feedback.
Moreover the degree of substructure is much higher for the simulated clusters with AGN feedback
and all clusters with AGN feedback were found to be disturbed.

This discrepancy in the substructure analysis between observations and simulations should be 
considered when refining simulation models, which may be affected both by the treatment of
the hydrodynamics and the sub-resolution galaxy formation physics. 
For example, the inclusion of AGN feedback in the simulations is necessary to reproduce 
the scaling relation of the observed X-ray clusters well, as was shown in~\cite{puchwein08} 
and in Fig.~\ref{fig:ep_so_lxt}. 
The substructure measures are, however, also sensitive to this physics.
They should thus be taken into account in addition to global cluster properties and ICM profiles. 
In our simulations, it might, for example, be interesting to explore if more accurate hydrodynamics, 
higher resolution that allows resolving AGN bubbles better, as well as a more accurate 
matching of the properties of these bubbles to observations would improve the agreement 
with observed substructure measures.

\begin{acknowledgements}
  HB and GC acknowledge support from the DFG Transregio Program TR33
  and the Munich Excellence Cluster ''Structure and Evolution of the Universe''.  
  GC acknowledges support by the DLR under grant no. 50 OR 1405.
  EP acknowledges support by the Kavli Foundation and the FP7 ERC Advanced Grant Emergence-320596.
\end{acknowledgements}

\footnotesize{
  \bibliographystyle{aa}
  \bibliography{subs}

\begin{thebibliography}{42}
\expandafter\ifx\csname natexlab\endcsname\relax\def\natexlab#1{#1}\fi

\bibitem[{{Allen} {et~al.}(2006){Allen}, {Dunn}, {Fabian}, {Taylor}, \&
  {Reynolds}}]{allen06}
{Allen}, S.~W., {Dunn}, R.~J.~H., {Fabian}, A.~C., {Taylor}, G.~B., \&
  {Reynolds}, C.~S. 2006, \mnras, 372, 21

\bibitem[{{Beck} {et~al.}(2016){Beck}, {Murante}, {Arth}, {Remus}, {Teklu},
  {Donnert}, {Planelles}, {Beck}, {F{\"o}rster}, {Imgrund}, {Dolag}, \&
  {Borgani}}]{beck16}
{Beck}, A.~M., {Murante}, G., {Arth}, A., {et~al.} 2016, \mnras, 455, 2110

\bibitem[{{B{\"o}hringer} {et~al.}(2015){B{\"o}hringer}, {Chon}, {Bristow}, \&
  {Collins}}]{r2_underdensity}
{B{\"o}hringer}, H., {Chon}, G., {Bristow}, M., \& {Collins}, C.~A. 2015, \aap,
  574, A26

\bibitem[{{B{\"o}hringer} {et~al.}(2014){B{\"o}hringer}, {Chon}, \&
  {Collins}}]{r2_cosm}
{B{\"o}hringer}, H., {Chon}, G., \& {Collins}, C.~A. 2014, \aap, 570, A31

\bibitem[{{B{\"o}hringer} {et~al.}(2013){B{\"o}hringer}, {Chon}, {Collins},
  {Guzzo}, {Nowak}, \& {Bobrovskyi}}]{reflex2-2}
{B{\"o}hringer}, H., {Chon}, G., {Collins}, C.~A., {et~al.} 2013, \aap, 555,
  A30

\bibitem[{{B{\"o}hringer} {et~al.}(2002){B{\"o}hringer}, {Matsushita},
  {Churazov}, {Ikebe}, \& {Chen}}]{boehringer02}
{B{\"o}hringer}, H., {Matsushita}, K., {Churazov}, E., {Ikebe}, Y., \& {Chen},
  Y. 2002, \aap, 382, 804

\bibitem[{{B{\"o}hringer} {et~al.}(2010){B{\"o}hringer}, {Pratt}, {Arnaud},
  {Borgani}, {Croston}, {Ponman}, {Ameglio}, {Temple}, \&
  {Dolag}}]{rexcess_sub}
{B{\"o}hringer}, H., {Pratt}, G.~W., {Arnaud}, M., {et~al.} 2010, \aap, 514,
  A32

\bibitem[{{B{\"o}hringer} {et~al.}(2001){B{\"o}hringer}, {Schuecker}, {Guzzo},
  {Collins}, {Voges}, {Schindler}, {Neumann}, {Cruddace}, {De Grandi},
  {Chincarini}, {Edge}, {MacGillivray}, \& {Shaver}}]{reflex1}
{B{\"o}hringer}, H., {Schuecker}, P., {Guzzo}, L., {et~al.} 2001, \aap, 369,
  826

\bibitem[{{B{\"o}hringer} {et~al.}(2007){B{\"o}hringer}, {Schuecker}, {Pratt},
  {Arnaud}, {Ponman}, {Croston}, {Borgani}, {Bower}, {Briel}, {Collins},
  {Donahue}, {Forman}, {Finoguenov}, {Geller}, {Guzzo}, {Henry}, {Kneissl},
  {Mohr}, {Matsushita}, {Mullis}, {Ohashi}, {Pedersen}, {Pierini}, {Quintana},
  {Raychaudhury}, {Reiprich}, {Romer}, {Rosati}, {Sabirli}, {Temple}, {Viana},
  {Vikhlinin}, {Voit}, \& {Zhang}}]{rexcess}
{B{\"o}hringer}, H., {Schuecker}, P., {Pratt}, G.~W., {et~al.} 2007, \aap, 469,
  363

\bibitem[{{Buote} \& {Tsai}(1995)}]{buote95}
{Buote}, D.~A. \& {Tsai}, J.~C. 1995, \apj, 452, 522

\bibitem[{{Chon} \& {B{\"o}hringer}(2012)}]{reflex2-1}
{Chon}, G. \& {B{\"o}hringer}, H. 2012, \aap, 538, A35

\bibitem[{{Chon} {et~al.}(2013){Chon}, {B{\"o}hringer}, \&
  {Nowak}}]{chon12_scl}
{Chon}, G., {B{\"o}hringer}, H., \& {Nowak}, N. 2013, \mnras, 429, 3272

\bibitem[{{Chon} {et~al.}(2012){Chon}, {B{\"o}hringer}, \& {Smith}}]{chon12}
{Chon}, G., {B{\"o}hringer}, H., \& {Smith}, G.~P. 2012, \aap, 548, A59

\bibitem[{{Collins} {et~al.}(2000){Collins}, {Guzzo}, {B{\"o}hringer},
  {Schuecker}, {Chincarini}, {Cruddace}, {De Grandi}, {MacGillivray},
  {Neumann}, {Schindler}, {Shaver}, \& {Voges}}]{collins00}
{Collins}, C.~A., {Guzzo}, L., {B{\"o}hringer}, H., {et~al.} 2000, \mnras, 319,
  939

\bibitem[{{Fabjan} {et~al.}(2010){Fabjan}, {Borgani}, {Tornatore}, {Saro},
  {Murante}, \& {Dolag}}]{fabjan10}
{Fabjan}, D., {Borgani}, S., {Tornatore}, L., {et~al.} 2010, \mnras, 401, 1670

\bibitem[{{Hahn} {et~al.}(2015){Hahn}, {Martizzi}, {Wu}, {Evrard}, {Teyssier},
  \& {Wechsler}}]{hahn15}
{Hahn}, O., {Martizzi}, D., {Wu}, H.-Y., {et~al.} 2015, ArXiv e-prints,
  arXiv:1509.04289

\bibitem[{{Hasselfield} {et~al.}(2013){Hasselfield}, {Hilton}, {Marriage},
  {Addison}, {Barrientos}, {Battaglia}, {Battistelli}, {Bond}, {Crichton},
  {Das}, {Devlin}, {Dicker}, {Dunkley}, {D{\"u}nner}, {Fowler}, {Gralla},
  {Hajian}, {Halpern}, {Hincks}, {Hlozek}, {Hughes}, {Infante}, {Irwin},
  {Kosowsky}, {Marsden}, {Menanteau}, {Moodley}, {Niemack}, {Nolta}, {Page},
  {Partridge}, {Reese}, {Schmitt}, {Sehgal}, {Sherwin}, {Sievers}, {Sif{\'o}n},
  {Spergel}, {Staggs}, {Swetz}, {Switzer}, {Thornton}, {Trac}, \&
  {Wollack}}]{hasselfield13}
{Hasselfield}, M., {Hilton}, M., {Marriage}, T.~A., {et~al.} 2013, \jcap, 7, 8

\bibitem[{{Heinz} {et~al.}(1998){Heinz}, {Reynolds}, \& {Begelman}}]{heinz98}
{Heinz}, S., {Reynolds}, C.~S., \& {Begelman}, M.~C. 1998, \apj, 501, 126

\bibitem[{{Kravtsov} \& {Borgani}(2012)}]{kravtsov12}
{Kravtsov}, A.~V. \& {Borgani}, S. 2012, \araa, 50, 353

\bibitem[{{Mahdavi} {et~al.}(2013){Mahdavi}, {Hoekstra}, {Babul}, {Bildfell},
  {Jeltema}, \& {Henry}}]{mahdavi13}
{Mahdavi}, A., {Hoekstra}, H., {Babul}, A., {et~al.} 2013, \apj, 767, 116

\bibitem[{{McCarthy} {et~al.}(2010){McCarthy}, {Schaye}, {Ponman}, {Bower},
  {Booth}, {Dalla Vecchia}, {Crain}, {Springel}, {Theuns}, \& {Wiersma}}]{mc10}
{McCarthy}, I.~G., {Schaye}, J., {Ponman}, T.~J., {et~al.} 2010, \mnras, 406,
  822

\bibitem[{{McNamara} \& {Nulsen}(2007)}]{mcnamara07}
{McNamara}, B.~R. \& {Nulsen}, P.~E.~J. 2007, \araa, 45, 117

\bibitem[{{McNamara} \& {Nulsen}(2012)}]{mcnamara12}
{McNamara}, B.~R. \& {Nulsen}, P.~E.~J. 2012, New Journal of Physics, 14,
  055023

\bibitem[{{Mead} {et~al.}(2010){Mead}, {King}, {Sijacki}, {Leonard},
  {Puchwein}, \& {McCarthy}}]{mead10}
{Mead}, J.~M.~G., {King}, L.~J., {Sijacki}, D., {et~al.} 2010, \mnras, 406, 434

\bibitem[{{Mohr} {et~al.}(1993){Mohr}, {Fabricant}, \& {Geller}}]{mohr93}
{Mohr}, J.~J., {Fabricant}, D.~G., \& {Geller}, M.~J. 1993, \apj, 413, 492

\bibitem[{{Peterson} {et~al.}(2001){Peterson}, {Paerels}, {Kaastra}, {Arnaud},
  {Reiprich}, {Fabian}, {Mushotzky}, {Jernigan}, \& {Sakelliou}}]{peterson01}
{Peterson}, J.~R., {Paerels}, F.~B.~S., {Kaastra}, J.~S., {et~al.} 2001, \aap,
  365, L104

\bibitem[{{Planck Collaboration} {et~al.}(2015{\natexlab{a}}){Planck
  Collaboration}, {Ade}, {Aghanim}, {Arnaud}, {Ashdown}, {Aumont},
  {Baccigalupi}, {Banday}, {Barreiro}, {Bartlett}, \& et~al.}]{planck15_cosm}
{Planck Collaboration}, {Ade}, P.~A.~R., {Aghanim}, N., {et~al.}
  2015{\natexlab{a}}, ArXiv e-prints, arXiv:1502.01589

\bibitem[{{Planck Collaboration} {et~al.}(2015{\natexlab{b}}){Planck
  Collaboration}, {Ade}, {Aghanim}, {Arnaud}, {Ashdown}, {Aumont},
  {Baccigalupi}, {Banday}, {Barreiro}, {Bartlett}, \&
  et~al.}]{planck15_cluster}
{Planck Collaboration}, {Ade}, P.~A.~R., {Aghanim}, N., {et~al.}
  2015{\natexlab{b}}, ArXiv e-prints, arXiv:1502.01597

\bibitem[{{Planelles} {et~al.}(2014){Planelles}, {Borgani}, {Fabjan},
  {Killedar}, {Murante}, {Granato}, {Ragone-Figueroa}, \& {Dolag}}]{plane14}
{Planelles}, S., {Borgani}, S., {Fabjan}, D., {et~al.} 2014, \mnras, 438, 195

\bibitem[{{Poole} {et~al.}(2006){Poole}, {Fardal}, {Babul}, {McCarthy},
  {Quinn}, \& {Wadsley}}]{poole06}
{Poole}, G.~B., {Fardal}, M.~A., {Babul}, A., {et~al.} 2006, \mnras, 373, 881

\bibitem[{{Pratt} {et~al.}(2009){Pratt}, {Croston}, {Arnaud}, \&
  {B{\"o}hringer}}]{pratt09}
{Pratt}, G.~W., {Croston}, J.~H., {Arnaud}, M., \& {B{\"o}hringer}, H. 2009,
  \aap, 498, 361

\bibitem[{{Puchwein} {et~al.}(2008){Puchwein}, {Sijacki}, \&
  {Springel}}]{puchwein08}
{Puchwein}, E., {Sijacki}, D., \& {Springel}, V. 2008, \apjl, 687, L53

\bibitem[{{Puchwein} {et~al.}(2010){Puchwein}, {Springel}, {Sijacki}, \&
  {Dolag}}]{puchwein10}
{Puchwein}, E., {Springel}, V., {Sijacki}, D., \& {Dolag}, K. 2010, \mnras,
  406, 936

\bibitem[{{Rasia} {et~al.}(2015){Rasia}, {Borgani}, {Murante}, {Planelles},
  {Beck}, {Biffi}, {Ragone-Figueroa}, {Granato}, {Steinborn}, \&
  {Dolag}}]{rasia15}
{Rasia}, E., {Borgani}, S., {Murante}, G., {et~al.} 2015, \apjl, 813, L17

\bibitem[{{Rasia} {et~al.}(2013){Rasia}, {Meneghetti}, \& {Ettori}}]{rasia13}
{Rasia}, E., {Meneghetti}, M., \& {Ettori}, S. 2013, The Astronomical Review,
  8, 40

\bibitem[{{Sembolini} {et~al.}(2016){Sembolini}, {Yepes}, {Pearce}, {Knebe},
  {Kay}, {Power}, {Cui}, {Beck}, {Borgani}, {Dalla Vecchia}, {Dav{\'e}},
  {Elahi}, {February}, {Huang}, {Hobbs}, {Katz}, {Lau}, {McCarthy}, {Murante},
  {Nagai}, {Nelson}, {Newton}, {Perret}, {Puchwein}, {Read}, {Saro}, {Schaye},
  {Teyssier}, \& {Thacker}}]{sembolini16}
{Sembolini}, F., {Yepes}, G., {Pearce}, F.~R., {et~al.} 2016, \mnras, 457, 4063

\bibitem[{{Sijacki} \& {Springel}(2006)}]{sijacki06}
{Sijacki}, D. \& {Springel}, V. 2006, \mnras, 366, 397

\bibitem[{{Sijacki} {et~al.}(2007){Sijacki}, {Springel}, {Di Matteo}, \&
  {Hernquist}}]{sijacki07}
{Sijacki}, D., {Springel}, V., {Di Matteo}, T., \& {Hernquist}, L. 2007,
  \mnras, 380, 877

\bibitem[{{Springel} {et~al.}(2005{\natexlab{a}}){Springel}, {Di Matteo}, \&
  {Hernquist}}]{springel05_bh}
{Springel}, V., {Di Matteo}, T., \& {Hernquist}, L. 2005{\natexlab{a}}, \mnras,
  361, 776

\bibitem[{{Springel} \& {Hernquist}(2002)}]{springel02}
{Springel}, V. \& {Hernquist}, L. 2002, \mnras, 333, 649

\bibitem[{{Springel} {et~al.}(2005{\natexlab{b}}){Springel}, {White},
  {Jenkins}, {Frenk}, {Yoshida}, {Gao}, {Navarro}, {Thacker}, {Croton},
  {Helly}, {Peacock}, {Cole}, {Thomas}, {Couchman}, {Evrard}, {Colberg}, \&
  {Pearce}}]{springel05}
{Springel}, V., {White}, S.~D.~M., {Jenkins}, A., {et~al.} 2005{\natexlab{b}},
  \nat, 435, 629

\bibitem[{{Vikhlinin} {et~al.}(2009){Vikhlinin}, {Kravtsov}, {Burenin},
  {Ebeling}, {Forman}, {Hornstrup}, {Jones}, {Murray}, {Nagai}, {Quintana}, \&
  {Voevodkin}}]{vik09}
{Vikhlinin}, A., {Kravtsov}, A.~V., {Burenin}, R.~A., {et~al.} 2009, \apj, 692,
  1060

\end{thebibliography}
}

\end{document}